\documentclass{aastex63}
\usepackage{amsmath}	
\usepackage{amssymb}
\usepackage{xcolor}

\received{***}
\revised{***}
\accepted{\today}
\submitjournal{ApJ}

\shorttitle{B0656+14 and the Monogem Ring}
\shortauthors{Jumei Yao et al.}

\begin{document}

\title{Interstellar scintillation and polarization of PSR~B0656+14 in the Monogem Ring}

\correspondingauthor{Jumei Yao}
\email{yaojumei@xao.ac.cn}
\author{Jumei Yao}
\affiliation{Xinjiang Astronomical Observatory, Chinese Academy of Sciences\\
150, Science 1-Street, Urumqi, Xinjiang 830011, China}
\author{William A. Coles}
\affiliation{Electrical and Computer Engineering, University of California,\\
San Diego, 92093, USA}
\author{Richard N. Manchester}
\affiliation{Australia Telescope National Facility, CSIRO Space and Astronomy \\
P.O. Box 76, Epping NSW 1710, Australia}
\author{Daniel R. Stinebring}
\affiliation{Department of Physics and Astronomy Oberlin College\\
Oberlin, OH 44074, USA} 
\author{Michael Kramer}
\affiliation{Max-Planck-Institut f\"ur Radioastronomie, Auf dem H\"ugel\\
69, D-53121 Bonn, Germany}
\affiliation{Jodrell Bank Centre for Astrophysics, The University of Manchester, M13 9PL, UK}
\author{Na Wang}
\affiliation{Xinjiang Astronomical Observatory, Chinese Academy of
  Sciences\\
  150, Science 1-Street, Urumqi, Xinjiang 830011, China}
\author{Di Li}
\affiliation{CAS Key Laboratory of FAST, National Astronomical Observatories, Chinese Academy of Sciences\\
Chaoyang District, Datun Road, A.20, Beijing 100101, China}
\affiliation{University of Chinese Academy of Sciences, Beijing 100049, China}
\affiliation{NAOC-UKZN Computational Astrophysics Centre, University of KwaZulu-Natal, Durban 4000, South Africa}
\author{Weiwei Zhu}
\affiliation{CAS Key Laboratory of FAST, National Astronomical Observatories, Chinese Academy of Sciences\\
Chaoyang District, Datun Road, A.20, Beijing 100101, China}
\author{Yi Feng}
\affiliation{Zhejiang Lab, Hangzhou, Zhejiang 311121, China}
\author{Jianping Yuan}
\affiliation{Xinjiang Astronomical Observatory, Chinese Academy of
  Sciences\\
  150, Science 1-Street, Urumqi, Xinjiang 830011, China}
\author{Pei Wang}
\affiliation{CAS Key Laboratory of FAST, National Astronomical Observatories, Chinese Academy of Sciences\\
Chaoyang District, Datun Road, A.20, Beijing 100101, China}



\begin{abstract}
High sensitivity interstellar scintillation and polarization observations of PSR~B0656+14 made at three epochs over a year using the Five-hundred-meter Aperture Spherical radio Telescope (FAST) show that the scattering is dominated by two different compact regions. We identify the one nearer to the pulsar with the shell of the Monogem Ring, thereby confirming the association. The other is probably associated with the Local Bubble. We find that the observed position angles of the pulsar spin axis and the space velocity are significantly different, with a separation of $19\fdg3\pm$0\fdg8, inconsistent with a previously published near-perfect alignment of $1\degr\pm 2\degr$. The two independent scattering regions are clearly defined in the secondary spectra which show two strong forward parabolic arcs. The arc curvatures imply that the scattering screens corresponding to the outer and inner arcs are located approximately 28~pc from PSR B0656+14 and 185~pc from the Earth, respectively. Comparison of the observed Doppler profiles with electromagnetic simulations shows that both scattering regions are mildly anisotropic. For the outer arc, we estimate the anisotropy $A_R$ to be approximately 1.3, with the scattering irregularities aligned parallel to the pulsar velocity. For the outer arc, we compare the observed delay profiles with delay profiles computed from a theoretical strong-scattering model. Our results suggest that the spatial spectrum of the scattering irregularities in the Monogem Ring is flatter than Kolmogorov, but further observations are required to confirm this.
\end{abstract}
 
\section{Introduction}
\label{sec:Intro}
The propagation of pulsar signals through the turbulent ionized interstellar medium (ISM) provides us with an opportunity to study the ISM using pulsar interstellar scintillation (ISS) observations. The two-dimensional (2D) autocovariance function (ACF) of the dynamic spectra can provide an estimate of the location of the scattering region and an estimate of its turbulence spectrum on very small scales ($\la$ 10,000 km) \citep[e.g.,][]{cr98, sss+03, sss06}. The 2D power spectrum of the dynamic spectrum, known as the secondary spectrum, often shows remarkable parabolic arcs \citep{smc+01}. When this is the case, the curvature of the arc can provide a more precise estimate of the location of the scattering region than the 2D ACF.

The parabolic arcs have two forms: a primary arc with its apex at the origin of the power spectrum; and an assembly of many reverse ``arclets" with their apexes distributed along the primary arc. The primary arc arises from interference of a lightly scattered wave from the pulsar with the angular spectrum of more heavily scattered waves. The curvature of these arcs depends on the location of the scattering region and the velocity of the pulsar. Consequently, arcs are only apparent if the scattering occurs in a compact region (or regions) along the line of sight to the pulsar. Such scattering sites have been found in the shell of Local Bubble \citep[e.g.,][]{bot+16, xlh+18}, in HII regions in the Galactic spiral arms \citep{fab+18} and in the shell of supernova remnants (SNRs)\citep{yzm+21}.  

To establish an association between a pulsar and a nearby SNR is often quite challenging. A reliable association should satisfy at least the first two criteria given by \citet{kas96}, similar distance and similar age. However, it is not an easy task to accurately measure independent distances and ages for either pulsars or SNRs. Further evidence for an association can come from  the detection of a pulsar proper motion vector pointing away from the SNR's center \citep[e.g.,][]{klh+03} or the direct interaction between a pulsar and the SNR \citep[e.g.,][]{sfs+89}. In \cite{yzm+21}, the ISS arc detection of PSR J0538+2817, which is located on the sky within the SNR S147 shows that the shell of this SNR dominates the scattering of  PSR J0538+2817 and provides us with a new method to further confirm the association between a pulsar and an SNR.

\begin{figure}
\center
	\includegraphics[width=10.0 cm, angle=0]{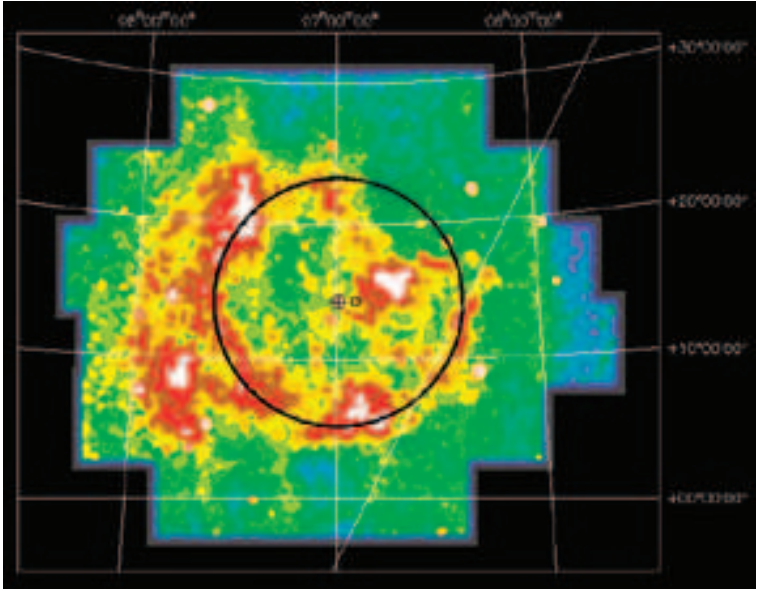}
    \caption{X-ray image of the Monogem Ring from the ROSAT all-sky survey in J2000 right ascension and declination \citep{tbb+03}. The current position of PSR B0656+14 is marked with a cross, and the Galactic plane is indicated by the sloping line west of the pulsar. A 9.2$^\circ$-radius circle centered on the pulsar position indicates the primary ring structure, and the estimated position of the pulsar 10$^5$~yrs ago is marked with a small square. The ring is imperfect -- there is an apparent blow-out to the east and a missing section to the north-west, perhaps due to foreground absorption or to slower expansion into a dense region. The small bright source in the north-west at 06:17 +22:34 is SNR IC443}.
    \label{fig:MR}
\end{figure}

After the ISS arc detection for PSR J0538+2817, we conducted FAST observations of five other pulsars located within SNRs, and PSR~B0656+14 is one of these. This 385-ms pulsar was discovered in the Second Molonglo pulsar survey \citep{mlt+78}, and later shown to be located close to the geometric center of the Monogem Ring \citep{chm+89, nch+81, tch+91}. The Monogem Ring is a $25\degr$ diameter ring of soft X-ray emission whose morphology shows significant deviations from circular symmetry as shown in Figure~\ref{fig:MR}. Besides the positional agreement, for PSR~B0656+14 the parallax distance of 288$^{+33}_{-27}$~pc \citep{btg+03} is consistent with the parallax distance of 282$^{+47}_{-34}$~pc for the star ``15 Mon" \citep{van07}, which is located in the southern region of the Monogem Ring \citep{psa+96,plk+97}, giving further evidence for the association. According to \citet{tbb+03}, the age of the Monogem Ring from Sedov modeling is 86~kyr, which is consistent with 110~kyr, the characteristic spin-down age of PSR~B0656+14. They also showed that the inferred birth position from the measured proper motion has an acceptable offset from the geometric center of the Monogem Ring. Together, these observations strongly support the association. 

For several young pulsars, including PSRs B0531+21, J0538+2817, B0833-45, B1706$-$44, B1951+32 and B0656+14, there is strong observational evidence for 2D alignment (in the plane of the sky) between the pulsar spin axis and proper motion \citep[e.g.,][]{nr04, jkk+07, wlj07}. Recently, using FAST observations of PSR J0538+2817, we found the first evidence for three-dimensional (3D) spin-velocity alignment in a pulsar \citep{yzm+21}. Through polarization analysis, we obtained the position angle (PA, measured from north toward east) of the spin axis $\psi_0$ and the inclination angle of the spin axis to the line of sight ($\zeta$), thereby establishing the 3D orientation of the spin axis. As SNR S147 is a near-perfect spherical shell, the ISS arc detection enabled us to estimate the radial velocity of PSR J0538+2817. This, combined with proper motion measurements, allowed us to establish the 3D orientation of the velocity vector. 

Unfortunately, the Monogem Ring is far from being a spherical shell, and so the ISS arc detection for PSR~B0656+14 does not give us information about its radial velocity. However, it helps us to confirm the association between the PSR~B0656+14 and the Monogem Ring and to prove that the SNR shell is dominating the pulsar scattering. PSR~B0656+14 has a measured proper motion of $\mu_{\alpha}=44.1\pm$0.6~mas~yr$^{-1}$ and $\mu_{\delta}=-2.4\pm$0.3~mas~yr$^{-1}$ \citep{btg+03}, giving a PA for the pulsar velocity of $\psi_{\rm pm}=93\fdg1 \pm 0\fdg4$. On the basis of Parkes polarization observations of PSR~B0656+14, \citet{jkk+07} found a PA of  $-86\degr$ (or $+94\degr$)~$\pm 2\degr$ for the projected spin axis. This puts PSR~B0656+14 in their Table~1 as one of the few pulsars with near-perfect 2D alignment ($\Delta$PA of $1\degr \pm 2\degr$) of the spin and velocity vectors.  

In this paper, we use high-sensitivity observations made with FAST \citep[see][for a detailed description of the telescope]{lwq+18} at frequencies around 1375~MHz to provide new evidence for the association of PSR~B0656+14 and the Monogem Ring. The secondary spectrum for PSR~B0656+14 shows two clear arcs and we use the curvature of these arcs to show that the scattering of the signal from PSR~B0656+14 occurs in the shell of the Monogem Ring (outer arc) and in shell of the Local Bubble (inner arc). A polarization analysis based on the FAST observations shows that the spin and velocity vectors for PSR~B0656+14 are significantly misaligned.

The arrangement of our paper is as follows: We describe the FAST observations, data processing procedures and ISS results in Section~\ref{sec:ISS}. We show the polarization results and discuss the 2D spin-velocity alignment in Section~\ref{sec:Pol}. In Section~\ref{sec:Discn}, we summarise our results and give our conclusions.

\section{Observations and interstellar scintillation of PSR~B0656+14}\label{sec:ISS}

In this section, we present ISS results for PSR~B0656+14 from three observations recorded at the FAST radio telescope using the central beam of the 19-beam receiver. We observed PSR~B0656+14 for 1~hr on MJD 59139 (2020 October 17) and 3~hrs on each of MJD 59183 (2020 November 30) and MJD 59512 (2021 October 25). The 19-beam receiver covers the frequency band 1050~MHz to 1450~MHz, but we only use the band 1300~MHz to 1450~MHz to avoid known radio frequency interference (RFI). We used the analysis program {\sc dspsr} \citep{sb11}\footnote{\url{http://dspsr.sourceforge.net}} and the {\sc psrchive} software package \citep{sdo12}\footnote{\url{http://psrchive.sourceforge.net}} to reduce our data. Data for each channel (bandwidth 0.122~MHz) were folded at the topocentric pulse period using a sub-integration time of 20~s and then polarization calibrated by using short observations of a pulsed noise source injected into the feed before the pulsar observation \citep[for details of the method see][]{str04}.

Following the data processing procedures described in \cite{yzm+21}, we obtained the 2D dynamic spectra (power versus radio frequency and time) for each of these observations using {\sc psrflux}. Our procedure for computing the secondary spectra is: First, we normalize the mean power of each sub-integration spectrum to the mean power across the observation to remove the effects of short-term pulse intensity variations; Second, we apply a Hamming window function to the outer 10\% of each dynamic spectrum to reduce the effects of aliasing in the secondary spectrum; Third, following \cite{chc+11}, to minimize spectral leakage we use first differences to pre-whiten the dynamic spectrum which typically has a steep spectrum since the spectrum of electron-density fluctuations in the interstellar medium is steep. Then, we form the secondary spectrum by using a 2D Fourier transform, take its squared magnitude and finally divide it by the square of a transfer function to recover our best estimate of the secondary spectrum (called post-darkening procedure). From \cite{chc+11}, the 2D transfer function is
\begin{equation}
H(f_t, f_\nu)=4\sin(2\pi f_t \Delta f)\sin(2\pi f_\nu \Delta t)
\end{equation}
where $\Delta t$ and $\Delta f$ are the sub-integration time and the channel bandwidth respectively, and $f_t$ and $f_\nu$ are the differential Doppler shift and the differential delay for each point in the secondary spectrum.

The primary objective of the ISS analysis based on the detected scintillation arcs in the secondary spectra is to determine the location of the scattering regions for comparison with interstellar features such as the Monogem Ring and the Local Bubble. A secondary objective is to determine the spectral exponent of the scattering turbulence and its anisotropy. To do these accurately, we use the following techniques: (a) we calibrate the curvature of the arcs in strong scattering using an electromagnetic simulation; (b) we include both the Earth's velocity and the screen velocity in estimating the scattering region location; and (c) we compare the Doppler profile and the delay profile with simulations and theory, respectively.

\subsection{Dynamic spectra and Autocovariance Functions}\label{sec:DS_ACF}
\begin{figure}
\center
	\includegraphics[width=7.0 cm, angle=270]{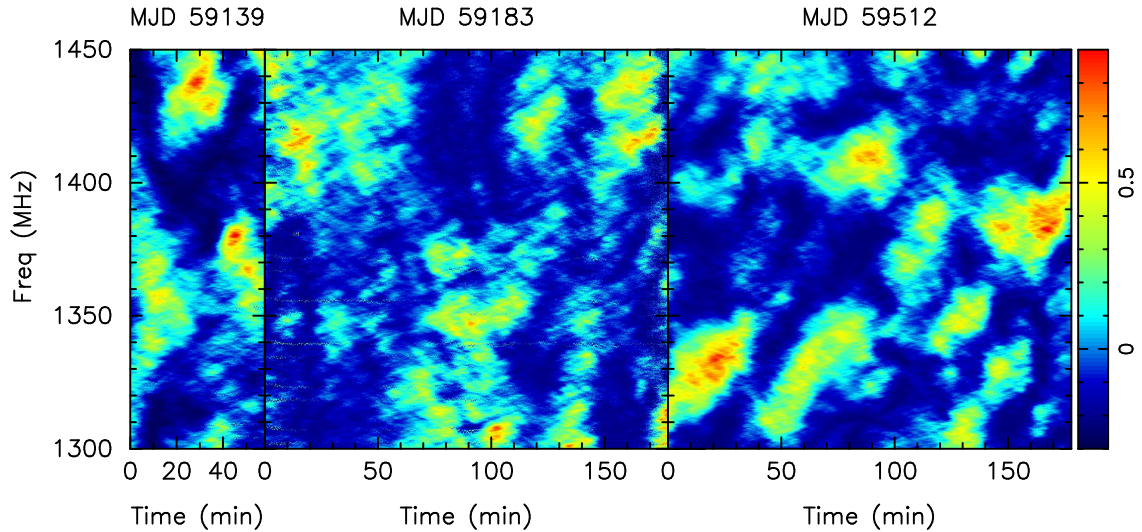}
    \caption{Dynamic spectra from FAST observation of PSR~B0656+14 for 150~MHz bands centered at 1375 MHz. These panels are spectra for the 1-h observations made on MJD 59139 (left) and 3-h observations made on MJDs 59183 (middle) and 59512 (right), respectively. The three panels are plotted with the same scales in both axes and the color scale is linear in signal power with arbitrary units.}
    \label{fig:ds}
\end{figure}
\begin{figure}
\center
	\includegraphics[width=9.0 cm, angle=270]{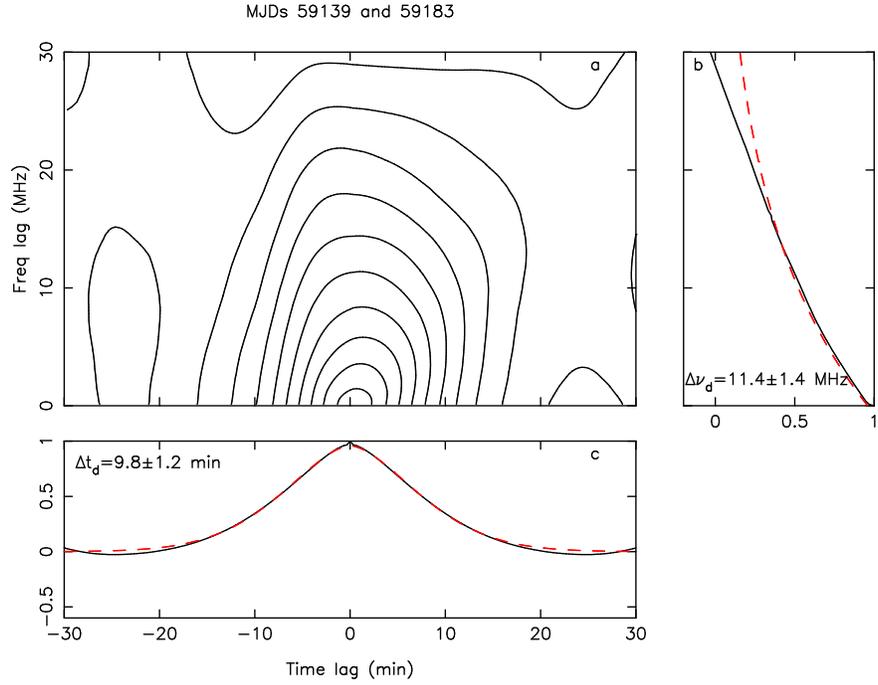}
    \caption{Average ACFs for MJDs 59139 and 59183 and the best-fit results for $\Delta \nu_d$ and $\Delta t_d$ at 1375~MHz. Panel a shows the averaged 2D ACF, panel b shows the corresponding 1D frequency-domain ACF with the red-dashed line being the fitted curve, and panel c shows the same for the 1D time-domain ACF.}
    \label{fig:acf}
\end{figure}

Figure~\ref{fig:ds} shows the dynamic spectra for PSR~B0656+14 from the three observations over the frequency band 1300~MHz to 1450~MHz. 
Following \cite{rch+19}, we determine the diffractive timescale, $\Delta t_d$, by doing a least-squares fit to the one-dimensional time-domain ACF for $\Delta t<10$~min using:
\begin{equation}\label{eq:scint_time1}
C(\Delta t, 0)=A\,\rm exp\left(-\left|\frac{\Delta t}{\Delta t_d}\right|^{\frac{5}{3}}\right)
\end{equation}
\begin{equation}\label{eq:scint_time2}
C(0,0)=A+W,
\end{equation}
where $W$ accounts for the uncorrelated noise in the data. We then fix $A$ and do a least-squares fit to the one-dimensional frequency-domain ACF for $\Delta \nu<20$~MHz using: 
\begin{equation}\label{eq:scint_freq}
C(0, \Delta\nu)=A\,\rm exp\left(-\left|\frac{\Delta \nu}{\Delta \nu_d/ln2}\right|\right).
\end{equation}
The 2D ACFs of the three observations show a skewness resulting from a phase gradient across the wavefront which differs but does not change sign across the three observations. Compared with the last observation, the first two show a much smaller phase gradient. To improve the signal-to-noise ratio (S/N), we average the ACFs for MJDs 59139 and 59183, giving the 2D ACF shown in Figure~\ref{fig:acf}, panel a. We then use Equations~\ref{eq:scint_time1} -- \ref{eq:scint_freq} to estimate $A=0.959\pm0.002$, W$=0.041\pm$0.002, $\Delta t_d=9.8\pm1.2$~min and $\Delta\nu_d=11.4\pm1.4$~MHz. The uncertainties of $\Delta t_d$ and $\Delta \nu_d$ include two parts, the statistical uncertainty from the data fitting and the fractional uncertainty from the finite number of observed scintles in the dynamic spectra \citep{wmj+05}. The 1D ACFs at zero lag and the best-fit results are shown in panels b and c of Figure~\ref{fig:acf}. 

\begin{figure}
\center
	\includegraphics[width=9.0 cm, angle=270]{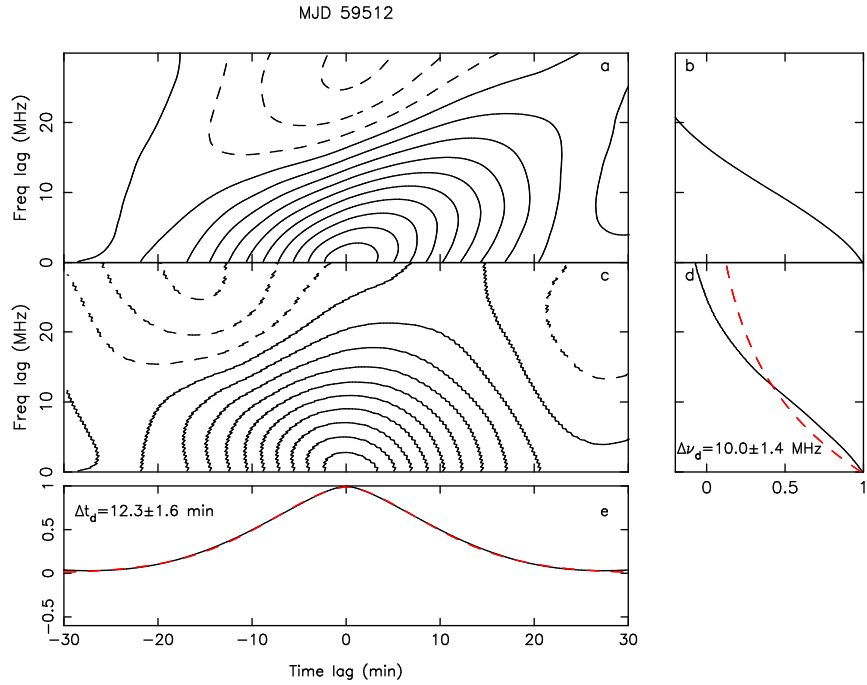}
    \caption{The ACF for MJD 59512 and the best-fit results for $\Delta \nu_d$ and $\Delta t_d$ at 1375~MHz. Panel a and c show the 2D ACF and the corrected 2D ACF, panel b and d show the corresponding 1D frequency-domain ACF, and panel e shows the 1D time-domain ACF. The red-dashed lines represent the best-fit results. }
    \label{fig:acf_59512}
\end{figure}

For MJD 59512, the phase gradient and hence the ACF skewness is much larger. Because of this we cannot use Equations~\ref{eq:scint_time1} -- \ref{eq:scint_freq} directly to estimate $\Delta \nu_d$. Instead, we use a model described in \citet{rcn+14} (Equation A6 of that paper) to correct for the refractive shift, selecting the best fit by eye. Figure~\ref{fig:acf_59512} shows the 2D ACF before correction (panel a) and after correction (panel c). We then fit for the ACF parameters using Equations~\ref{eq:scint_time1} -- \ref{eq:scint_freq} with the corrected ACF, giving  $A=0.987\pm0.002$, W$=0.013\pm$0.002, $\Delta t_d=12.3\pm1.6$~min and $\Delta\nu_d=10.0\pm1.4$~MHz. Within the uncertainties, the frequency and lag widths, $\Delta t_d$ and $\Delta\nu_d$, are consistent with the results obtained from the average ACF for MJDs 59139 and 59183.

Based on 1380~MHz observations from Westerbork Synthesis Radio Telescope, \cite{wwsr06} estimate a  scintillation time scale for PSR~B0656+14 of order 1000~s and a scintillation bandwidth of $\la 80$~MHz. They also note that the scintillation bandwidth of their 327~MHz Arecibo observations must be comparable to or less than their individual channel bandwidth of 0.098~MHz. All three of these estimates are consistent with our observations.

The scintillation strength, defined as $u=\sqrt{\nu/\nu_d}$ \citep{ric90}, is $11.0\pm 0.7$, indicating that at 1375~MHz this pulsar is out of the weak scintillation regime but not in the asymptotically strong regime. Some refractive scintillation might be observed \citep[cf.,][Fig. 20]{crg+10}. Importantly, as discussed below,
the scattered electric field for an ideal thin screen can be calculated directly from the wave equation. Thus by simulating a turbulent screen with a given spatial power spectrum of phase, we can calculate the electric field as if it were observed. We can then duplicate the analysis process exactly, obtaining the simulated secondary spectrum without any assumptions about the brightness distribution.

Although the scintillation is not weak, the number of ``scintles", i.e. degrees of freedom in the combined ACFs, is only $\sim$ 500. This is insufficient to estimate the phase structure function reliably from the 1D temporal ACF, as was done by \citep{yzm+21} for PSR~J0538+2817.

\subsection{Secondary spectra and arc curvature}\label{sec:SS}
\begin{figure}
\center
	\includegraphics[width=12.0 cm, angle=270]{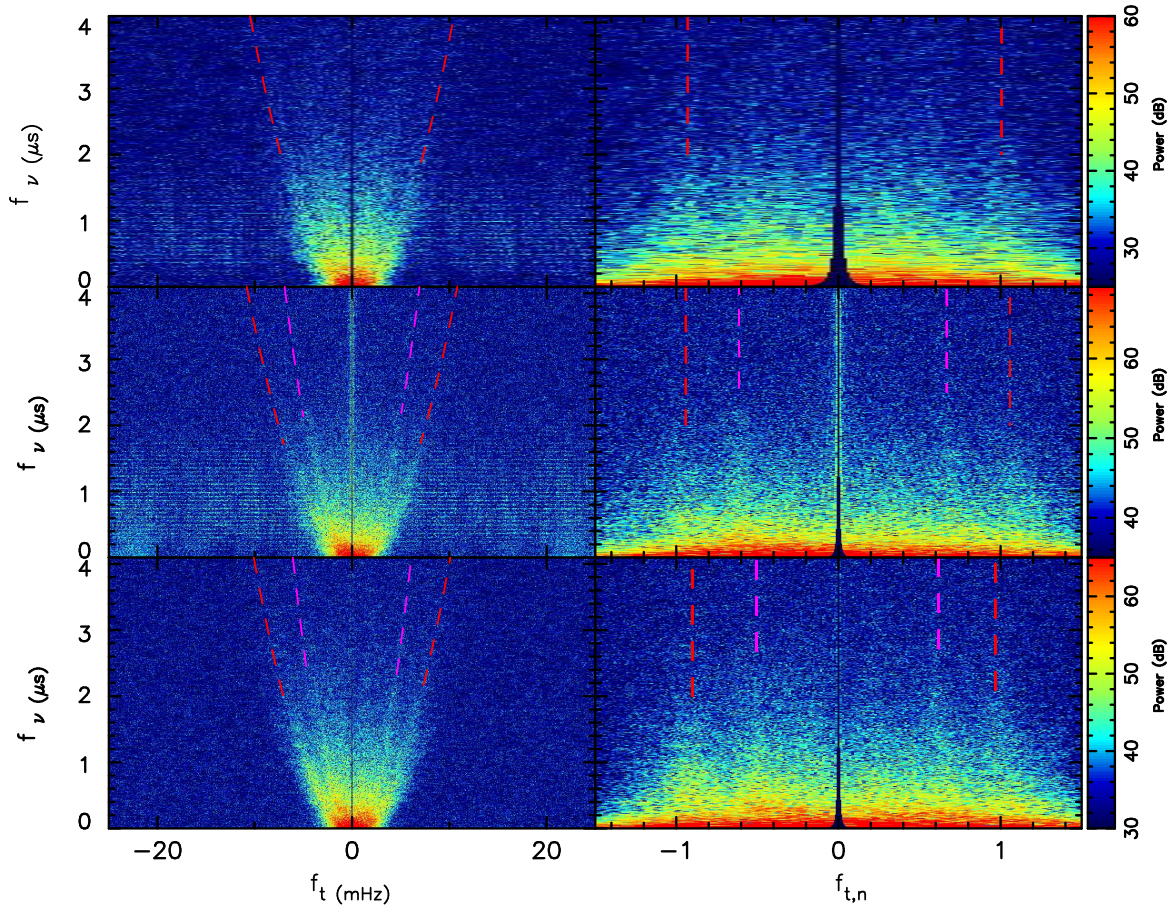}
    \caption{Secondary spectra (left panels) and normalized secondary spectra (right panels) for PSR~B0656+14 from observations made on MJDs 59139(top), 59183 (middle) and 59512 (bottom). The color scale is logarithmic in dB. In the left panels, the red dashed lines represent the central position of the outer arc for each epoch. For MJDs 59139 and 59512, the central position of an inner arc is shown in magenta. In the right panels, the red and magenta dashed lines show the best-fit position at negative and positive Doppler shifts for the outer arc and the inner arc,  respectively}
    \label{fig:ss_nor}
\end{figure}

For each of the dynamic spectra shown in Figure~\ref{fig:ds}, the corresponding  secondary spectra are shown in Figure~\ref{fig:ss_nor}. The Nyquist frequencies corresponding to the channel bandwidth and sub-integration time are $f_t(\rm {Nyquist}) = 25$~mHz and $f_\nu(\rm {Nyquist}) = 4.1$~$\mu$s, respectively. The first two observations show horizontal striations in the secondary spectra that are probably caused by a standing-wave problem in the FAST system. For the 1-hr observation on MJD 59139, there is an outer parabolic arc with diffuse inner structure, whereas the 3-hr observations on MJDs 59183 and 59512 also show a clear inner arc.

\begin{figure}
\center
	\includegraphics[width=12.0 cm, angle=270]{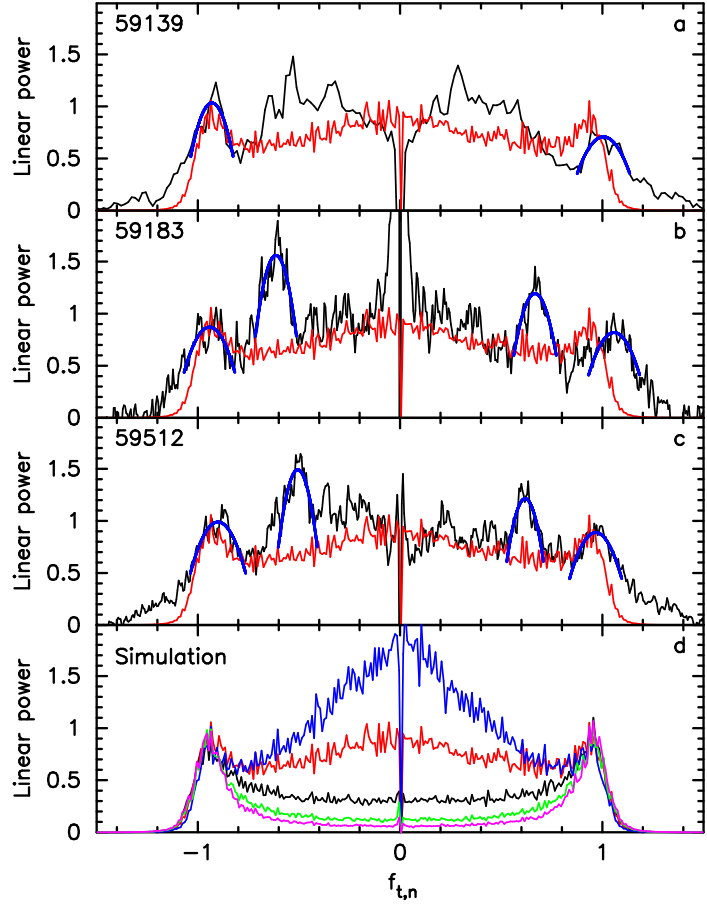}
    \caption{Observed and simulated Doppler profiles. Observed Doppler profiles are shown as black lines in panels a, b and c and blue lines show the parabolic fits to the arc peaks. Simulated Doppler profiles for the outer arcs are shown in all panels. For panels a, b and c the optimal case is shown in red, whereas panel d shows the simulated profiles for different degrees of anisotropy and orientation of the scattered image relative to the pulsar velocity across the screen. The different cases are as follows: black -- isotropic; green -- anisotropic with the major axis of the scattered image aligned with velocity with $A_R=1.3$; magenta -- aligned with $A_R=1.6$; red --  anisotropic with the major axis of the scattered image perpendicular to velocity with $A_R=1.3$; blue -- perpendicular with $A_R=1.6$.}
    \label{fig:fitting}
\end{figure}

We describe parabolic arcs using
\begin{equation}
f_\nu=\eta f^2_t
\end{equation}
where $\eta$ is the arc curvature \citep{crs+06}. To estimate the arc curvature for both the inner ($\eta_{\rm inner}$) and outer ($\eta_{\rm outer}$) arcs, following \cite{rcb+20}, we transformed the secondary spectra with respect to an arc curvature of $\eta_0=0.04$~s$^3$, such that an arc with curvature $\eta_0/\beta^2$ becomes a vertical line at ``normalised" $f_{t,n} = \beta$. The normalized secondary spectra are shown in the right panels of Figure~\ref{fig:ss_nor}. The normalization process also has the useful effect of reducing the standing-wave striations in the normalized spectra. To estimate the arc curvature optimally, we average the power along the $f_\nu$ axis to form the power distribution as a function of $f_{t,n}$, the so-called ``Doppler profile". To reduce the effect of strong signals at small $f_\nu$, we average the power only for $f_\nu > 1.0$~$\mu$s. We calculate the noise level in the normalized spectrum by averaging regions with 1.5$<\mid{f_{t,n}}\mid<$2.0 and subtract this mean noise level from the entire profile. Then, we normalize the $f_{t,n}$ axis so that the mean value of the peaks of the outer arc is 1.0. Figure~\ref{fig:fitting} shows the observed Doppler profiles for the three epochs in panels a, b and c.

As illustrated in Figure~\ref{fig:ss_nor}, the normalizing procedure over-sampled the secondary spectra, especially for regions with smaller $f_\nu$. Consequently, the Doppler profiles are also over-sampled and it is difficult to find reasonable regions to fit each peak. We first smooth the observed Doppler profiles using the 1-D Gaussian filter Gaussian1DKernel from {\sc Astropy }\footnote{\url{https://docs.astropy.org/en/stable/api/astropy.convolution.Gaussian1DKernel.html}} \citep{aps+18}. Since the MJD 59183 and 59512 observations are over 3~hr, compared to 1~hr for MJD 59139, the Doppler profile for the MJD 59139 observation has a factor of three lower resolution. However, we choose to smooth all three normalised Doppler profiles to the same resolution, viz., 0.08. 
We then use a parabolic function to fit a suitable region for each peak in the smoothed Doppler profiles to obtain the best estimate of $\beta$. Because of the smoothing, these estimates will underestimate the uncertainties. Consequently, we fit the selected peak regions in the raw Doppler profiles using the smoothed estimate as our initial guess. The resultant fits are shown in panels a, b and c of Figure~\ref{fig:fitting}. For each arc, we obtain the value of $\beta$ for the negative and positive Doppler shifts separately and designate these as $\beta_{N}$ and $\beta_{P}$. These are listed in Table 1 for the three epochs. The uncertainties in $\beta_{N}$ and $\beta_{P}$ are 1-$\sigma$ values given by the fitting algorithm. We verified these estimates using a bootstrap (with replacement) analysis with 100 iterations. To do this, we took the residuals between the data and the fitted parabola, shuffled them and then added them back to the fitted curve. The 1-$\sigma$ widths of the bootstrap distributions agreed well with the tabulated uncertainties.
As the secondary spectrum for MJD 59139 has a diffuse inner arc, we only obtain the arc curvature of the outer arc in that case.

Because of the modulation resulting from the phase gradient, the arcs shift slightly \citep{crg+10}, resulting in a small difference between the values of $|\beta_{N}|$ and $\beta_{P}$. As the phase gradient is small, we tested the effect of simply averaging $|\beta_{N}|$ and $\beta_{P}$ to obtain the correct curvature. We found this provided good compensation for offsets $\sim$10\% such as we observe. We therefore give the mean value, $\beta_M$, in Table~\ref{tab:arc_curvature}.

\begin{table}
    \centering
    \caption{Value and 1-$\sigma$ uncertainties of $\beta$ values and arc curvatures $\eta$ for both the inner and outer arcs.}
    \begin{tabular}{cccccc}
    	\hline
	MJD & $\beta_{N}$ &$\beta_{P}$ &  $\beta_{M}$ & $\beta_{MC}$& $\eta$\\
    &  &  & & & (s$^3$) \\
    \hline
        59139 & $-$0.9300$\pm$0.0059& 1.0054$\pm$0.0061& 0.9677$\pm$0.0042&1.0350$\pm$0.0045 &0.03734$\pm$ 0.00033\\
        59183 & $-$0.9422$\pm$0.0055& 1.0579$\pm$0.0059&  1.0001$\pm$0.0040&1.0696$\pm$0.0043&0.03497$\pm$0.00028\\          
        & $-$0.6138$\pm$0.0028&0.6667$\pm$0.0027&0.6403$\pm$0.0019& 0.6848$\pm$0.0021&0.08531$\pm$0.00052\\ 
        59512 & $-$0.9000$\pm$0.0034 & 0.9667$\pm$0.0026& 0.9334$\pm$0.0021& 0.9982$\pm$0.0023&0.04014$\pm$0.00018\\          
        & $-$0.5056$\pm$0.0045 &0.6180$\pm$0.0039& 0.5618$\pm$0.0030&0.6009$\pm$0.0032&0.11080$\pm$0.00117\\
    \hline
    \end{tabular}
    \label{tab:arc_curvature}
\end{table}

The widely used formulas for arc curvature \citep[e.g.,][]{crs+06}, are only valid for forward arcs in weak scintillation. In that case very sharp arcs form due to interference between the unscattered wave and the surrounding angular spectrum of scattered waves. Forward arcs persist into strong scattering if the scattering is not too anisotropic (axial ratio $A_R \la 3$) because there is a central core of lightly scattered waves which interferes with the surrounding highly scattered angular spectrum. Each element of this core produces an arc but each arc has a slightly different curvature and apex so the summation of arcs results in a broader arc centered at a smaller Doppler value (higher curvature). A theoretical analysis of this mechanism is possible for very strong scattering where the elements of the scattered angular spectrum are independent, but our observations are not in this regime. Consequently, we simulate the dynamic spectrum directly. We use an electromagnetic code which computes the electric field in the observing plane for a thin random phase screen. The phase screen is simulated to model a single realization of a given spatial power spectrum of electron density \citep{crg+10}. From this we can calculate the simulated dynamic spectrum of intensity and its secondary spectrum exactly as we would for an observation. 
The lowest panel of Figure~\ref{fig:fitting} shows a family of such simulations with different anisotropies, in both direction and degree, all with Kolmogorov spectra and the same strength of scintillation as the actual observations. The Doppler profile is normalized such that an ideal forward arc would appear at $\pm$1. The simulated arcs all peak near $|f_{t,n}| = 0.935$ which gives a curvature 13\% larger than ideal, so we correct all the measured arcs using $\beta_{MC} = \beta_{M}/0.935$ and calculate $\eta$ using $\beta_{MC}$. These values are given in Table~\ref{tab:arc_curvature}.
The uncertainties in $\beta_M, \beta_{MC}\, {\rm and}\, \eta$ are derived from those in $\beta_N\, {\rm and}\, \beta_P$ by error propagation.

We have not attempted to fit the observations directly because there is no theory or simulation applicable to multiple arcs in strong scintillation (in weak scintillation, superposition holds so multiple arcs are easily dealt with). However, many observational examples show multiple forward arcs in strong scattering and they appear to superimpose independently. We assume this to be the case and choose the simulation of the outer arc that just fits below the inner arcs. It is clear from Figure~\ref{fig:fitting} that the amplitude of the outer arcs and the power in their interior regions are well fitted by an anisotropy $A_R = 1.3$ in the scattered image with its major axis perpendicular to the velocity (plotted in red). This implies that the major axis of the scattering irregularities is parallel to the pulsar velocity. However, the inner arcs in panels b and c cannot have a central power excess. Consequently, the scattered image for these must be either isotropic or with an anisotropy oriented perpendicular to that of the outer arc.

\subsection{Locating the scattering regions}\label{sec:LSR}
To obtain the location of the scattering regions (defined by s and D) we again follow \citet{crs+06}.
The secondary spectrum is $S(f_\nu,f_t)$, where the differential delay is 
$f_\nu=[D(1-s)|\vec{\theta}|^2] /2cs$ and the differential Doppler shift is
$f_t=(\nu_c \vec{V}_{\rm eff, \perp} \cdot \vec{\theta})/cs$.
Here, $\vec{\theta} = (\theta_\alpha,\theta_\delta)$ is the scattering angle, $\vec{V}_{\rm eff,\perp}$ is the velocity of the line of sight through the scattering region, the scattering screen is located at a distance $sD$ from the pulsar, $\nu_c$ is the band-center frequency (1375~MHz), and all vectors are in the plane of the sky $(\alpha,\delta)$. Based on the parallax measurement \citep{btg+03}, we adopt a distance for the pulsar of $D = 290\pm30$~pc. The boundary arc is defined by $f_\nu$ and the maximum $f_t$ for that delay. Hence:
\begin{align}
(f_\nu, f_t)_{\rm arc} &= [{D(1-s)}\,\theta^2 /2cs,\,{\nu_c V_{\rm eff,\perp}}\,\theta/cs], \label{eq:max_ft} \,\,\, {\rm where} \\
\vec{V}_{\rm eff,\perp} &= (1-s) \vec{V}_{\rm pulsar,\perp}  + s \vec{V}_{{\rm Earth, \perp}} - \vec{V}_{\rm scr, \perp} \label{eq:v_eff}.
\end{align}
From these equations we obtain the curvature: 
\begin{equation}
\eta=4625\frac{D_{\rm kpc}\,s(1-s)}{\nu^2_{\rm GHz}|V_{\rm eff,\perp}|^2}\label{eq:arc_c},   
\end{equation}
where $V_{\rm eff,\perp}$ is in km~s$^{-1}$ and $\eta$ is in $s^{3}$. The curvature does not depend on $\vec{\theta}$ or any of the scattering characteristics of the plasma. This makes it immune to the variations in level of turbulence that plague measurements of $\Delta t_d$ and $\Delta \nu_d$.

To estimate $s$ for each arc at each epoch we must solve Equation~\ref{eq:arc_c} which is quadratic in $s$. We have estimates of all the parameters necessary to solve for $s$ given $\eta$, except for $V_{\rm scr,\perp}$. We use the CALCEPH Library \citep{glmf15} to obtain the Earth's velocity components which vary with the orbital phase of the Earth. We expect $V_{\rm scr,\perp}$ to be small compared with ${V}_{\rm pulsar,\perp}$, so we first assume $V_{\rm scr,\perp} = 0$. For the outer arc, one of the two solutions for $s$ at each epoch is very close to the Earth and hence unlikely.  Our best estimates are therefore $s = 0.1668\pm 0.0012$, $0.1763 \pm 0.0015$ and $0.1817\pm 0.0008$ for MJDs 59139, 59183 and 59512, respectively. These imply that the scattering screen responsible for the outer arc is very close to the shell of the Monogem ring. Accordingly we use the known velocity and geometry of the Monogem ring to estimate $V_{\rm scr,\perp}$ for the outer arc. In contrast, for the inner arc, the near-Earth solutions at both epochs are feasible astronomically. If we had only one observation we would not have been confident of either the near-pulsar or the near-Earth solutions for the inner arc. However, with two epochs, only the near-pulsar solutions agree. The solutions for the inner arc are therefore $s = 0.4289\pm 0.0026$ and $0.4140\pm 0.0029$. This screen may be associated with the Local Bubble but we are unable to use the association to estimate $V_{\rm scr,\perp}$ for this arc.

If the pulsar is at the center of the expanding shell of the ring, then $V_{\rm scr, \perp} = 0$ for the outer arc. However, the X-ray image of the Monogem Ring shown in Figure~\ref{fig:MR} suggests that the pulsar has moved a substantial distance since it was born. Following the discussion in \citet{tbb+03}, from the pulsar characteristic age, the time since birth of the pulsar and the SNR is about $10^5$~yr. Given the pulsar transverse velocity $V_{\rm pulsar,\perp}$ of about $60\;{\rm km\,s}^{-1}$, it would have moved $\theta_c \sim 1\fdg2$ in right ascension and a few arc minutes in declination since birth. Measurements of optical absorption lines suggest that the current expansion velocity of the shell, $V_{\rm expn}\sim 105~{\rm km\,s}^{-1}$ \citep{jss+98}. From \citet{tbb+03}, the angular radius of the primary shell $\theta_s$ is about $9\fdg2$. Hence, $V_{\rm scr,\alpha} \sim (\theta_c/\theta_s) V_{\rm expn} \sim 14\;$km\,s$^{-1}$. However \citet{tbb+03} note that there is an uncertainty in the location of the center of the shell of order $1^\circ$. This results in an uncertainty of $\sim 10\;\rm km\,s^{-1}$ in $V_{\rm scr, \alpha}$.

We solved for $s_{\rm outer}$ again with $V_{\rm scr, \alpha} = 14\;\rm{km\,s}^{-1}$ which gives $s_{\rm outer}$ = 0.0994 $\pm$ 0.0007, 0.1017 $\pm$ 0.0008 and 0.1076 $\pm$ 0.0004 for MJDs 59139, 59183 and 59512. These are still inside the Monogem ring and confirm the association, but the values for the three epochs are significantly different. We found that no constant $V_{\rm scr,\alpha}$ within $10\;{\rm km\,s}^{-1}$ of $V_{\rm scr, \alpha} = 14\;{\rm km\,s}^{-1}$ will fit all three observations within the expected uncertainty in $s_{\rm outer}$. However we can fit all three observations, matching $s_{\rm outer}$ exactly for the mean $V_{\rm scr, \alpha}$ in the range $14\pm10\;{\rm km\,s}^{-1}$, provided that we allow $V_{\rm scr, \alpha}$ to increase by approximately 1.6\;$\rm km\,s^{-1}$ from MJD 59139 to 59512. This velocity increase probably results from the trajectory of the line-of-sight through the scattering medium due to the pulsar proper motion. Further observations might help to clarify this.

The values of $s_{\rm inner}$ at MJD 59183 and 59512 do not agree either, and we searched for a value of $V_{\rm scr, \alpha}$ for the Local Bubble that would improve the match. We found that the two values of $s_{\rm inner}$ were close to equal at $V_{\rm src, \alpha}=4\;\rm km\,s^{-1}$ and they remained within 3$\sigma$ of the uncertainty of the difference for $V_{\rm scr, \alpha}$ in the range 0 to 10$\;\rm km\,s^{-1}$.
We also found that the observations on MJD 59183 are particularly sensitive to $V_{\rm scr, \alpha} < 0$. There is no solution at all for $V_{\rm scr, \alpha} < -6\, {\rm kms}^{-1}$. Accordingly we take the most probable value of $V_{\rm src, \alpha}$ to be $4\;\rm km\,s^{-1}$ and assume that it lies in the range [0,10]$\;\rm km\,s^{-1}$.

\begin{table}
\caption{Value and uncertainty of $s$, $D_{\rm ps}$ and $D_{\rm es}$ obtained from MC analyses for four different cases. $D_{\rm ps}$ is given for the outer arc and $D_{\rm es}$ for the inner arc. $D_{\rm ps}$ and $D_{\rm es}$ are in pc.}
\centering
\setlength\tabcolsep{3.0pt}
\footnotesize
\begin{tabular}{cccccccccccc}
	\hline
	   MJD &\multicolumn{2}{c}{$\eta$ varied}& &\multicolumn{2}{c}{$\eta$, $\mu_{\alpha}$, $\mu_{\delta}$ varied}&
		&\multicolumn{2}{c}{$\eta$, $\mu_{\alpha}$, $\mu_{\delta}$, $D$ varied} & &\multicolumn{2}{c}{$\eta$, $\mu_{\alpha}$, $\mu_{\delta}$, $D$, $V_{\rm scr, \perp}$ varied}\\
		 & s &$D_{\rm ps}$ or $D_{\rm es}$& &s& $D_{\rm ps}$ or $D_{\rm es}$& &s&$D_{\rm ps}$ or $D_{\rm es}$& & s& $D_{\rm ps}$ or $ D_{\rm es}$\\
		\hline
		  59139&0.0994$\pm$0.0007& 28.82$\pm$0.22& & 0.0994$\pm$0.0032&28.83$\pm$0.93& & 0.100$\pm$0.015& 28.8$\pm7.4$ & &0.097$\pm$0.043& 28$\pm13$\\
		59183&0.1017$\pm$0.0008& 29.49$\pm$0.22& & 0.1017$\pm$0.0035&29.50$\pm$0.99& & 0.102$\pm$0.016& 29.5$\pm7.6$ & &0.098$\pm$0.044& 28$\pm14$\\
		 &0.3585$\pm$0.0022& 186.04$\pm$0.63& & 0.3585$\pm$0.0087& 186.0$\pm2.5$&
		 & 0.359$\pm$0.030& 186$\pm12$& &0.355$\pm$0.047& 186$\pm16$\\
		59512&0.1076$\pm$0.0004& 31.21$\pm$0.12& & 0.1077$\pm$0.0034&31.22$\pm$0.97& & 0.108$\pm$0.017& 31.2$\pm8.0$ & &0.104$\pm$0.046& 30$\pm15$\\
		&0.3577$\pm$0.0027& 186.27$\pm$0.79& & 0.3577$\pm$0.0076&186.3$\pm$2.2& &0.358$\pm$0.029&186$\pm$12& & 0.355$\pm$0.042&187$\pm$15\\
		\hline
	\end{tabular}
	\label{tab:MC_cases}
\end{table}

\begin{figure}
\center
	\includegraphics[width=9.0 cm, angle=270]{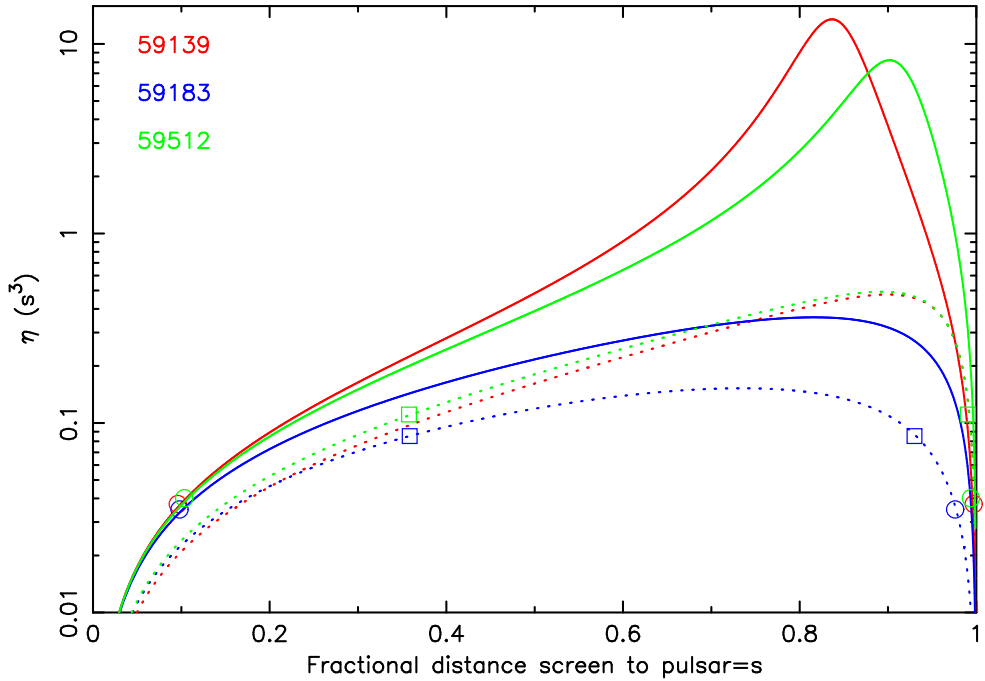}
    \caption{Arc curvature $\eta$ versus $s$ derived using Equation~\ref{eq:arc_c} for MJDs 59139, 59183 and 59512. The full and dotted lines are calculated for the outer arc with $V_{\rm scr,\alpha}=14~\rm km\,s^{-1}$ and the inner arc with $V_{\rm scr,\alpha} = 4~\rm km\,s^{-1}$, respectively. The circles and squares show the measured $\eta$ and corresponding values of $s$ for the outer and inner arcs, respectively. For the near-Earth solutions of the outer arc, the right side circles, we adopt $V_{\rm scr,\alpha} = 4~\rm km\,s^{-1}$.}
    \label{fig:eta_s}
\end{figure}

The solutions for the locations of the scattering regions are shown in Figure~\ref{fig:eta_s} and tabulated in Table~\ref{tab:MC_cases}. As the solutions are very non-linear, we use a set of Monte Carlo (MC) simulations to obtain the uncertainties. In each case we use the mean parameters described above, but we perform different MC variations in each case. For the first case we vary only $\eta$ according to the observational uncertainties in Table 1. This allows us to compare the estimates at different epochs and judge the goodness of fit of a constant $V_{\rm scr, \perp}$ model. For the second case we vary $\eta$ as before but we also vary $\mu$. For the third case we add variation in $D$. The MC variations in all these parameters are generated with a gaussian distribution. However the fourth case, in which we add MC variations to $V_{\rm scr, \perp}$, we have to use a distribution with finite support for the MC analysis of $s_{\rm inner}$ to keep the extreme values of the gaussian distribution from the divergent region of the solution space for MJD 59183. We use a beta distribution $\beta$(2.0,2.6)$\times 10$ on the interval (0, 10)~km/s peaked at about 4~km/s for the inner arc. The outer arc is not as sensitive and we use a gaussian distribution for it.

It is clear that the effect of the uncertainties increases strongly from cases 1 to 4. Based on the fourth case in Table~\ref{tab:MC_cases}, we obtain  best estimates of the location of the two scattering regions as follows:
\begin{align}
s_{\rm inner} &= 0.360\pm0.045,  & s_{\rm outer} &= 0.099\pm0.044, \label{eq:mean_s} \\
D_{\rm es, inner} &= 185\pm15~{\rm pc}
 \hspace*{2cm} \rm{and} & D_{\rm ps, outer} &= 28\pm14~{\rm pc}.\label{eq:d_ps}
 \end{align}

There is little doubt that the scattering causing the outer arc is associated with the Monogem Ring. However, it is of interest to investigate the actual location of the scattering screen. As Table~\ref{tab:MC_cases} shows, the uncertainty in $D$ is a much larger contributor to our estimate of $D_{\rm ps, outer}$ than the uncertainties in $\eta$ and $(\mu_{\alpha},\mu_{\delta})$. To partially bypass the large uncertainty in $D$, we use the relative screen distance $D_{\rm ps, outer}/D = s_{\rm outer}$. If we assume that the scattering is occurring on a spherical shell, for the nominal value $s_{\rm outer} = 0.1$, 
the implied angular radius of this shell is $\theta_{R,s_{\rm outer}} = 5\fdg71$. This is substantially less than the nominal radius of the Shell, viz. $10\degr$ or even $12\fdg5$ \citep{tbb+03}. Even taking the upper limit for $s_{\rm outer}$ of about 0.145 (Equation~\ref{eq:mean_s}), $\theta_{R,s_{\rm outer}} \sim 8\fdg2$, still less than the nominal shell radius. Possible explanations are: a) for a spherical shell, that the scattering region is well within the shell; or b) the shell is non-spherical and is closer to the pulsar on the near side with the scattering is occurring close to the shell boundary. Of course, a spherical shell is a very nominal representation of the actual structure of the Monogem Ring in 2D, let alone in 3D. Consequently, it is quite possible that the ring and an associated scattering region is closer to the pulsar on the near side. These results illustrate the importance of improving the pulsar distance estimate, and to a lesser extent, the proper motion estimate. For example, with two or three more observations at different phases of the Earth's orbit we should be able to quantify the screen perpendicular velocity and its gradient.

The X-ray image of \citet{tbb+03} shows a large feature extending about $5\degr$ to the west of the main shell. It is unlikely that this feature is part of a shell extending in front of the pulsar as we would have seen evidence for it in the secondary spectra. 

For the inner arc, the distance between the scattering screen and the Earth is $D(1-s_{\rm inner})$, i.e., $185\pm 15$~pc. Some local pulsars are scattered by a screen located 100 -- 200~pc from the Earth, and the shell of the Local Bubble has long been suspected as the compact ionized region dominating the scattering \citep[see, e.g.,][]{bot+14, xlh+18}. As measured by \cite{pfb+20} (see their Figure 1), the boundary of the Local Bubble in direction of PSR~B0656+14 is 160 -- 240~pc from the Earth, consistent with our inner arc resulting from scattering in the shell of this Bubble.

We can learn more from the power in the Doppler profiles at small $|f_{t,n}|$ from Figure~\ref{fig:eta_s}. For example, at MJD 59183 the maximum possible $\eta$ for $V_{\rm scr,\alpha} =4\;\rm km\,s^{-1}$ is about 0.11~s$^3$. This corresponds to a minimum $|\beta| = 0.7$, i.e., just inside the inner arc. Therefore, none of the power in the Doppler profile for MJD 59183 inside the inner arc, including the broad spike at the origin, can be due to unresolved forward arcs.

Using the method described in \cite{yzm+21}, we used the measured $\Delta \nu_d$, $\Delta t_d$, s and $D_{\rm ps}$ for the outer arc of PSR~B0656+14, and a screen velocity of $14\pm 10\;\rm km\,s^{-1}$, to derive an axial ratio of the scattered image of $A_R=1.8\pm0.6$, with the major axis perpendicular to the pulsar velocity.
It is clear from Figure~\ref{fig:fitting} that $A_R$ cannot exceed 1.3 because contributions from the inner arcs must be positive definite. Therefore the Doppler profile of the outer arc alone must lie below any inner arcs. Our estimate of $A_R$ = 1.3 based on the Doppler profiles is consistent with the result derived from the measured $\Delta \nu_d$ and $\Delta t_d$, but is clearly a better estimate.

\subsection{Spatial Spectra of micro-turbulence}\label{sec:SST}
\begin{figure}
\center
	\includegraphics[width=11.0 cm, angle=270]{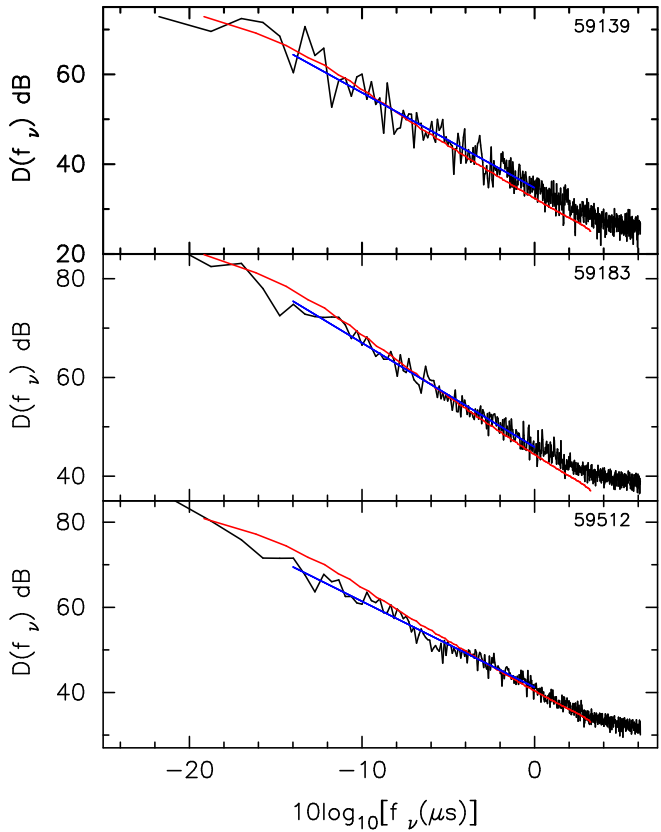}
    \caption{Observed and modelled delay profiles at 1375 MHz for MJDs 59139, 59183 and 59512. The black lines are the observed delay profiles, the red lines represent the modelled delay profiles for a Kolmogorov spectrum, and the blue lines show the best-fit slope to the linear part, i.e., delay lags between 0.04~$\mu$s and 1.00~$\mu$s. 
    }
    \label{fig:spectral_exponent}
\end{figure}

\cite{yzm+21} used the well-estimated temporal ACF$(t,0)$ to estimate the spectral exponent of the scattering turbulence for PSR~J0538+2817. However that is not possible with our observations of PSR~B0656+14 because there are not enough ``scintles" in the dynamic spectra to provide the necessary accuracy for ACF$(t,0)$. However the same information, to even smaller spatial scales, can be obtained from the delay profile. To form the delay profiles, we summed over the outer arc regions marked with blue lines in Figure~\ref{fig:fitting} (panels a -- c). The resulting delay profiles are shown as black lines in Figure~\ref{fig:spectral_exponent} for each epoch. They have a dynamic range in power of $\sim 50$~dB which cannot be achieved in single-precision simulations. However, delay profiles can be computed from the strong scintillation ACF model given by \cite{rcn+14} \citep[cf.,][]{lr99}. 
We Fourier transformed this model ACF to obtain the secondary spectrum and then calculated the delay profile as above. These model profiles, after matching to the observed delay amplitude, are shown in each panel of Figure~\ref{fig:spectral_exponent}. The model ACF from \cite{rcn+14} and its corresponding secondary spectrum are universal for asymptotically strong scintillation. The delay profile can be shown to converge to a power-law at large lags and we used that to correct for aliasing near the Nyquist frequency. For a Kolmogorov fluctuation spectrum, which we assume, the asymptotic exponent is $(-11/3 -1)/2 = -2.333$. The model delay profile tends to flatten at lags $\la 0.04$~$\mu$s. For these small lags, the amplitude of the scattering perturbations are relatively large and the relation between screen spatial frequency and scattering angle becomes non-linear \citep{lr99}. 

To estimate the delay profile exponent, we limited the delay range to 0.04~$\mu$s -- 1.00~$\mu$s to avoid the curvature at smaller delays and the white noise contribution at longer delays. The fits to the selected regions result in exponents of $-$2.12$\pm$0.07, $-$2.11$\pm$0.04 and $-2.02\pm0.04$ for MJDs 59139, 59183 and 59512, respectively. 
These imply spatial power spectral exponents of $-$3.24$\pm$0.14, $-$3.22$\pm$0.08 and $-3.04\pm0.08$, respectively. These exponents are significantly flatter than the Kolmogorov value of $-$3.67. Although the formal errors on the exponent are the same for MJDs 59183 and 59512, we have more confidence in the MJD 59512 value because its white noise spectrum is only 35\% as strong as that of MJD 59183. We determine this from the ACF fits, which provide both signal and white noise variance. The improved white noise is likely because the standing-wave problem in the receiver was corrected.
However the spectral exponent should be confirmed with additional observational data. If confirmed, it would be the first evidence for non-Kolmogorov behavior at the smallest spatial scales in the magneto-hydrodynamic process, perhaps shedding light on the dissipation mechanism.

For the outer arcs, based on Equation~\ref{eq:max_ft}, the values of $\theta$ corresponding to the delay limits of 0.04~$\mu$s and 1.00~$\mu$s are $0.16\pm0.01$~mas and $0.79\pm0.07$~mas, respectively. Using Equation~2.4 of \citet{ric90}, we find that corresponding spatial scales in the scattering plasma range range from 3.6$\times10^5$~km to 1.8$\times10^6$~km, a dynamic range of five.

\section{Polarization of PSR~B0656+14}\label{sec:Pol}
PSR B0656+14 is an interesting pulsar in its own right and has been widely observed. Here we discuss our FAST observations of dispersion measure (DM), rotation measure (RM), and the position angle (PA) of the linear polarization, and we put these in the context of earlier observations. We find that both DM and RM vary by more than their statistical uncertainty and we discuss these variations in detail. We use the observed PA to estimate the orientation of the spin axis and we compare it with the direction of the proper motion.


\subsection{Time Variability of RM and DM}\label{sec:DM_RM}
In order to measure the pulsar DM we used a standard timing analysis over the entire bandwidth 1050 -- 1450 MHz after manually minimizing the effects of RFI. The results are shown in fifth column of the upper part of Table~\ref{tab:RM}. For estimating the RM we use data for the two bands 1050 -- 1150~MHz and 1350 -- 1450~MHz, which are less affected by RFI.
For the RM we use  the program {\sc rmfit} \citep{sdo12}.  First, we sum the Stokes parameters across all channels of the two bands and search for a maximum in the fractional linear polarization of the pulse profile over the RM range of $\pm$1500~rad~m$^{-2}$. Then, we use {\sc rmfit} to iteratively refine the initial guess as follows: for each of the two bands we integrate the Stokes parameters across the band as a function of pulse phase, and then compute a weighted differential PA 
between the two bands. This gives an improved RM estimate. The process is then repeated until the change in RM between iterations is less than a preset threshold. The best-fit RMs (RM$_{\rm obs}$) are given in the second column of the upper part of Table~\ref{tab:RM} where each hour of the 3-hr observations on MJDs 59183 and 59512 has been separately analysed.

For accurate comparisons of RM$_{\rm obs}$, it is necessary to estimate the ionospheric component RM$_{\rm iono}$ because it is much more variable than the uncertainty on 
RM$_{\rm obs}$.
We used the routine {\sc ionFR} \citep{ssh+13} and the values of the ionospheric electron column density from the NASA CDDIS GNSS website.\footnote{\url{https://cddis.nasa.gov/archive/gnss/products/ionex/}} The derived RM$_{\rm iono}$ and the corrected RM$_{\rm ism}$ values are shown in the third and fourth columns of Table~\ref{tab:RM}. 

In the lower part of Table~\ref{tab:RM} we give four measurements of DM and two measurements of RM at earlier epochs from the literature. 
The Parkes observations of MJD 53663 \citep{jkk+07} were not corrected for the ionospheric component, but as RM$_{\rm iono}$ generally lies in the range $-5$\;rad~m$^{-2} < \rm{RM}_{\rm iono} < -1$\;rad~m$^{-2}$, they provide a useful lower bound.

\begin{table}
\caption{Observed DM, RM and the ionospheric and ISM contributions to it.}\label{tab:RM}
\centering
\begin{tabular}{cccccc}
\hline
MJD and UT & RM$_{\rm obs}$ & RM$_{\rm iono}$ & RM$_{\rm ism}$ & DM  \\
& (rad.~m$^{-2}$) & (rad.~m$^{-2}$) & (rad.~m$^{-2}$) & pc cm$^{-3}$ \\
\hline
59139 (UT 22~h) &   +23.05 $\pm$ 0.12   &   +0.59 $\pm$ 0.08  & +22.46 $\pm$ 0.14 & 13.959 $\pm$ 0.011 \\
59183 (UT 18~h) &   +23.57 $\pm$ 0.18   &   +0.90 $\pm$ 0.06  & +22.67 $\pm$ 0.19 &   \\
59183 (UT 19~h) &   +23.58 $\pm$ 0.18   &   +0.91 $\pm$ 0.07  & +22.67 $\pm$ 0.19 &  13.892 $\pm$ 0.018 \\
59183 (UT 20~h) &   +23.45 $\pm$ 0.15   &   +0.84 $\pm$ 0.04  & +22.61 $\pm$ 0.16 &   \\
59512 (UT 19~h) &   +23.85 $\pm$ 0.11   &   +0.97 $\pm$ 0.08  & +22.88 $\pm$ 0.14 &  \\
59512 (UT 20~h) &   +23.80 $\pm$ 0.14   &   +0.87 $\pm$ 0.07  & +22.93 $\pm$ 0.16  & 13.932 $\pm$ 0.012  \\
59512 (UT 21~h) &   +23.97 $\pm$ 0.10   &   +0.83 $\pm$ 0.07  & +23.14 $\pm$ 0.12 &  \\
\hline
48423$^a$ &&&& 14.02 \\
49721$^b$ &&&&13.977 $\pm$ 0.013 \\
53663$^c$ &+23.0$\pm$0.3& $-5\lesssim{\rm RM}_{\rm iono}\lesssim-1$& $+24\lesssim{\rm RM}_{\rm ism}\lesssim+28$ & 13.66$\pm$0.22  \\
56747.75$^d$&+28.00 $\pm$ 0.02&+5.26 $\pm$ 0.07&+22.73 $\pm$ 0.08& 14.076 $\pm$ 0.002 \\
\hline

\end{tabular}
\\
$^a$\cite{hr10}, $^b$\cite{hlk+04}, $^c$\cite{jkk+07}, $^d$\cite{sbg+19}.\\
\end{table} 

Our recent measurements of RM$_{\rm ism}$ show a linear increase of $\sim$0.5~rad~m$^{-2}$ over the year of observations, about three times the uncertainty. This raises the question of whether this is a real increase, a statistical fluctuation, or a result of under-estimating the uncertainty. We consider this question in Appendix B and conclude that it is likely to result from real changes in electron density in the Monogem Ring on a spatial scale of $\sim$12 AU.

\subsection{Rotating-vector model}\label{sec:RVM}

\begin{figure}
\center
    \includegraphics[width=12.0 cm, angle=270]{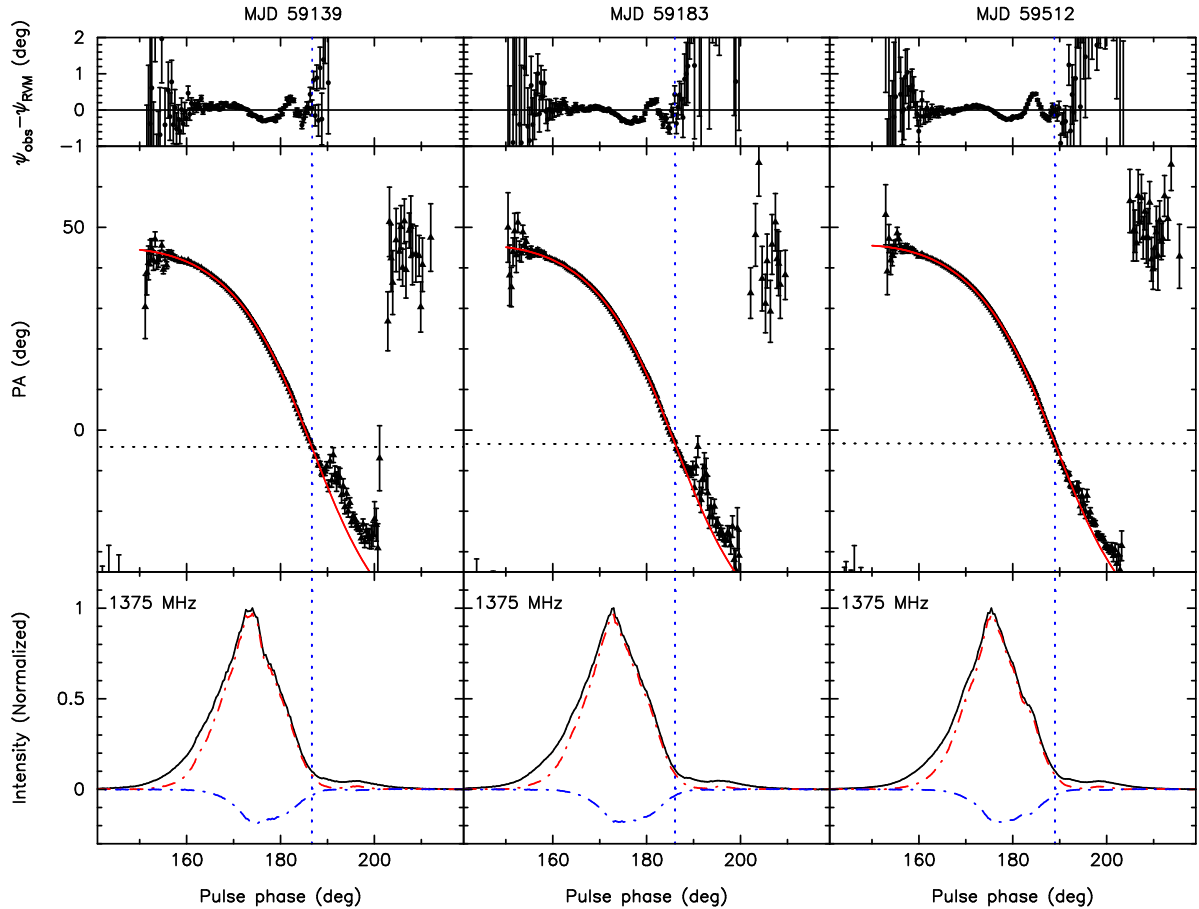}
    \caption{Polarization profiles for PSR~B0656+14 at 1375~MHz from observations made at MJDs 59139 (left), 59183 (middle) and 59512 (right). The bottom panels show the total intensity (solid black lines), the linear polarization (red dash-dot lines), and circular polarization (blue dash-dot lines). The middle panels show the observed PAs ($\psi$) at 1375~MHz as a function of pulse phase, and the red lines give the best-fit RVM solution. The top panels show the fit residuals. The vertical dotted lines show the central pulse phase from the RVM fit, $\phi_0$, and the horizontal dotted lines give the corresponding PA at 1375~MHz, $\psi_0$. }
    \label{fig:pa}
\end{figure}

The polarization profiles for PSR~B0656+14 are shown in Figure~\ref{fig:pa}. The observed PAs, defined as increasing counter-clockwise on the sky, and the circular polarization, follow the PSR/IEEE conventions \citep[see, e.g.,][]{ew01,smj+10,ksv+21}. A rotating vector model (RVM) \citep{rc69a}, was fitted to the observed PAs and is shown in red over the observed PAs in the middle panels. The PA residuals are shown in the top panels. The fitting of an RVM model to PA observations is a notoriously difficult problem \citep{ew01}. Our analysis, which was complex, is discussed in Appendix~\ref{app:rvm}. One can see in the middle panels that the overall fit is very good, but in the residuals it is apparent that systematic variations are much larger than the measurement errors and these dominate the uncertainty in the best fit parameters of the model.

The RVM for a simple dipole is described by  equation~\ref{eq:rvm}. Here $\psi$ is the PA and $\phi$ is the pulse phase. At the closest approach of the line of sight to the magnetic axis $\phi = \phi_0$ and $\psi = \psi_0$. The angle between the spin axis and the magnetic axis is $\alpha$ and $\zeta$ is the inclination angle of the spin axis from the line of sight. The impact parameter is $\beta = \zeta - \alpha$.
\begin{equation}
\psi=\psi_0 + \arctan \left [ \frac{\sin\alpha\, \sin(\phi-\phi_0)}{\sin\zeta\,\cos\alpha-
\cos\zeta\,\sin\alpha\,\cos(\phi-\phi_0)}\right ],
\label{eq:rvm}
\end{equation}

As described in Appendix B, we were able to fit the full RVM model to the MJD 59512 data by defining a maximum slope parameter MXG = $\sin(\alpha)/\sin(\beta)$ and fitting for $\alpha$, MXG, $\phi_0$ and $\psi_0$. However we could not fit all 4 parameters to either of the earlier epochs. Since we are primarily interested in the orientation of the spin axis $\psi_0$, we fit the earlier epochs only with $\phi_0$ and $\psi_0$, holding MXG and $\alpha$ at the values found for MJD 59512. This means that the uncertainties for $\psi_0$ and $\phi_0$ on the earlier epochs are misleading, but the mean values are useful. The results are given in Table~\ref{tab:RVM}.

\begin{table}
\caption{RVM best-fit results, mean values of RM$_{\rm obs}$, and the corresponding values of intrinsic $\psi_0$ at the three epochs. We use the PSR/IEEE sign convention for these angles.}\label{tab:RVM}
\centering
\begin{tabular}{ccccccc}
	\hline
	MJD & $\alpha$ & sin$\alpha$/sin$\beta$&$\psi_0$ & $\phi_0$
	RM$_{\rm obs}$ & $\psi_0$(intrinsic)\\
    & (deg.) &  & (deg.) & (deg.) & (rad.~m$^{-2}$) & (deg.)\\
    \hline
    59139 & 168 & $-2.98$& $-4.15$ & 186.66 & +23.05$\pm$0.12 & -66.9\\ 
    59183 & 168  &$-2.98$ &$-3.41$ & 186.05 & +23.53$\pm$0.10 & $-67.5$\\ 
    59512 &   168$\pm$9 & $-2.98\pm$0.02&$-3.32\pm$0.61 & 188.99$\pm$0.20& +23.87$\pm$0.07 & $-68.3\pm0.6$\\ 
	\hline
	\end{tabular}
\end{table} 

In order to compare $\psi_0$ with other determinations of the PAs of the projected spin axis and pulsar velocity, we must correct it to infinite frequency to give the so-called ``intrinsic" PA of the spin axis. We use the mean RM$_{\rm obs}$ for each observation, given in column~6 of Table~\ref{tab:RVM}, for this correction. The RMS of these three estimates is 0.7 deg, which is very close to the estimated uncertainty on MJD 59512, of 0.6 deg. Accordingly we take the mean of these three as a best estimate for the intrinsic $\phi_0$ = -67.6 (or +112.4) $\pm$ 0.7 deg.

This is $19\fdg3\pm 0\fdg8$ from the position angle of the pulsar transverse velocity ($\psi_{\rm pm}=93\fdg1\pm 0\fdg4$). Based on Parkes observations at 0.69~GHz and 3.1~GHz, \cite{jkk+07} estimated that $\psi_0 {\rm (intrinsic)}=-86\degr \pm 2\degr$\;(or $+94\degr \pm 2\degr$). This is just $1\degr\pm 2\degr$ from $\psi_{\rm pm}$, i.e.,  near-perfect 2D alignment, which is quite different from our result.

\subsection{Pulsar spin-velocity alignment}\label{section:P_V}
\begin{figure}
\center
 \includegraphics[width=6.0 cm, angle=270]{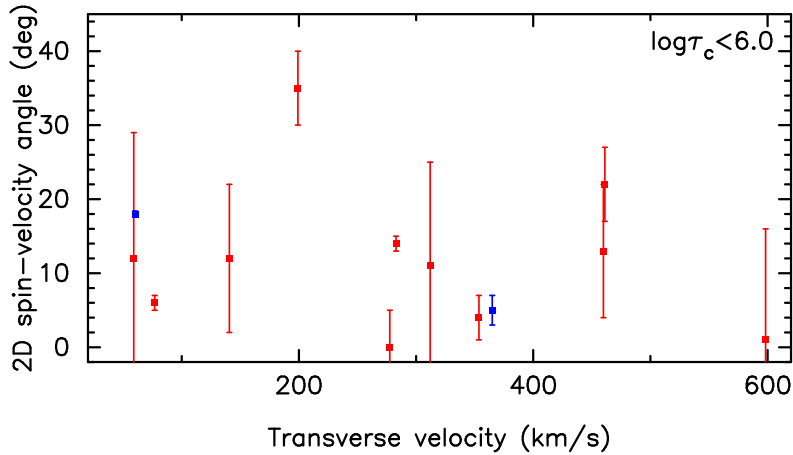}
    \caption{The 2D spin-velocity misalignment angle as a function of pulsar transverse velocity. These pulsars all have characteristic age less than $10^6$~yr and independent distance measurements. Blue: PSRs J0538+2817 and B0656+14 which have FAST  polarization data; Red: pulsars from Table~1 of \citet{nkc+12}.}  \label{fig:V_angle}
\end{figure}

After the publication of the first observed 3D spin-velocity alignment, for PSR J0538+2817 \citep{yzm+21}, \citet{jwk+22} proposed a novel explanation for pulsar spin-velocity alignment, i.e., that the displacement of the pulsar from the explosion center due to the initial kick channeled the direction of the fallback matter, thereby resulting in pulsar spin-velocity alignment. In this model, pulsars with larger space velocity tend to have smaller spin-velocity misalignment angle. Compared with PSR J0538+2817, which has a transverse velocity of 365$\pm$52~km/s, PSR~B0656+14 has a lower velocity of 60 $\pm$7~km/s and a larger 2D spin-velocity angle (blue points in Figure~\ref{fig:V_angle}) , which is consistent with the \citet{jwk+22} model. 

To check this with a larger sample, we selected other pulsars with characteristic age $\tau_c<6.0\times 10^6$~yr and with independent distances from Table~1 of \citet{nkc+12}. As Figure~\ref{fig:V_angle} shows, only two previous measurements have an uncertainty comparable to the FAST measurements. Never-the-less we include the less precise measurements to illustrate the current situation, viz., that there are too few high-precision measurements available to see the trend of spin-velocity misalignment angle with pulsar transverse velocity. High-sensitivity FAST polarization data for more pulsars will be very helpful for checking this relationship in the future.

\section{Summary and Conclusions}\label{sec:Discn}
We have used high sensitivity FAST observations to study the scintillation and polarization of PSR~B0656+14 and its relationship with the Monogem Ring. Secondary spectra for observations on MJDs 59139, 59183 and 59512 in late 2020 showed scintillation arcs, with two clear arcs seen in the second and third observations which are more sensitive. The mean curvature for the outer arc shows that the scattering screen is located a fractional distance of 0.099 $\pm$ 0.044 from the pulsar. Given the complex structure of the ring and the uncertainty in the birth location, this is consistent with a scattering screen close to or within the shell of the ring, and confirms the association between the PSR~B0656+14 and the Monogem Ring. It provides further evidence that the shell of an old SNR can dominate the scattering along the path to a pulsar \citep{yzm+21}. The curvature of the inner arc shows that, for this arc, the scattering screen is located 185$\pm$15~pc from the Earth, consistent with a location at the boundary of the Local Bubble. Observations of other, mostly local, pulsars by \citet{bgr98} and others have suggested that this boundary has the small-scale structure in its electron-density distribution needed to cause such scintillation features. Based on simulations of the Doppler profiles for a range of anisotropies, we found that the observations are well described by a scattering region for the outer arc which has an anisotropy $A_R \sim 1.3$ with the scattering irregularities aligned parallel to the pulsar velocity. The simulations also suggest that the scattering irregularities for the inner arc are either isotropic or aligned perpendicular to those for the outer arc. 
Modelling of the delay profiles suggests that all three observations are consistent with a turbulence spectrum for the scattering irregularities near the Monogem Ring which is slightly flatter than Kolmogorov.

Our observations of DM and RM show time variations exceeding the measurement uncertainty. Comparison with earlier measurement of DM and RM also shows significant variation. After reconsidering the estimated uncertainty, we conclude that both the electron density and the magnetic field in the scattering region near the Monogem Ring vary on a spatial scale of $\sim$12 AU.

At frequencies around 1~GHz, PSR~B0656+14 is nearly 100\% linearly polarized with a PA swing across the main pulse component that is well fitted by the RVM. 
The RVM fit together with the measured RM gives a projected direction for the pulsar spin axis on the plane of the sky of $\psi_0 = 112\fdg4\pm 0\fdg7$. This indicates a misalignment of approximately $19\degr$ between the projected direction of the pulsar spin axis and that of the space velocity ($\psi_{\rm pm}=93\fdg1\pm 0\fdg4$), contrary to a previous report suggesting close alignment of the two vectors. 

Given FAST's high sensitivity, we can expect to greatly increase the sample of pulsars with ISS arc detections and to improve the accuracy of many pulsar polarization measurements.  New ISS arc detections give us the opportunity to investigate the kind of ionized compact structures that dominate pulsar scattering and perhaps also to confirm some other pulsar -- SNR associations.

\acknowledgments
This work was supported by the National Natural Science Foundation of China Grant No. 11903049, 12041304, U2031117 and West Light Foundation of the Chinese Academy of Sciences (No. 2018-XBQNXZ- B-023). JMY was supported by Cultivation Project for FAST Scientific Payoff and Research Achievement of CAMS-CAS. DRS acknowledges support from the US National Science Foundation through Grant No. 2009759.

\begin{appendix}
\section{Variations in ionospheric RM and interstellar RM}\label{app:rm}
As discussed in Section~\ref{sec:DM_RM}, we derive apparently significant variations of RM$_{\rm ism}$ and DM across the year spanned by our FAST observations. In this Appendix we detail how these quantities and their uncertainties are measured as 
there are three possible explanations for the variations in RM$_{\rm ism}$.

The first is simply random fluctuations in the values of RM$_{\rm obs}$. The variations are about three times their estimated uncertainty. Such deviations are not uncommon in astronomy and generally result from unrecognised systematic errors. However the variations in RM$_{\rm ism}$ are correlated and appear linear. The significance of the linear slope is 3.7 sigma.

The second possibility is that the uncertainties in RM$_{\rm iono}$ are under-estimated. The ionospheric electron column density values from GNSS are given at intervals of $5\degr$ in Galactic longitude and $2\fdg5$ in Galactic latitude, and at 2-hourly intervals. As discussed by \citet{pnt+19}, because of possible small-scale and short-term variations there are significant uncertainties in the interpolation to the actual observatory location and observation time. Furthermore, there are large variations in the effective height of the ionospheric layer, typically by about a factor of two, from 300~km to 600~km, on daily, yearly and solar-cycle timescales which affect the calculation of the integrated electron content along the slant path to the pulsar. These uncertainties are especially large when the elevation angle of the pulsar at the observatory is small, i.e., when it is rising or setting. However the ionospheric error is a relatively small component of the total error. It would have to increase 75\% to increase the total error by 25\%.

The third possibility is that there is a real variation of RM$_{\rm ism}$ on the timescale of our observations, i.e., about one year. If the RM variations that we observe are real, they probably occur in the shell of the Monogem Ring. The transverse velocity due to proper motion of the pulsar is $V_{\rm pulsar,\perp} \sim 60$~km~s$^{-1}$ with an annual variation of $\pm 30$~km~s$^{-1}$ due to the Earth's orbital motion. Over the one year spanned by our observations, the line of sight to the pulsar in the shell would have moved by $\sim 12$~AU. Over 12 AU the fluctuation $\Delta$RM $= 0.8\,\Delta$DM\,B$_{||} + 0.8\, \Delta$B$_{||}$\,DM, where B$_{||}$ is the line-of-sight component of the interstellar magnetic field. We have no way to estimate $\Delta$\,B$_{||}$ but we can estimate the first term using the structure function of DM, $D_{\rm DM}(s) = \langle ({\rm DM}(r+s) - {\rm DM}(r))\rangle$. Here the expectation is denoted by $\langle \rangle$ and it is taken to be stationary with respect to transverse position $r$.

X-ray observations \citep{psa+96} suggest compression ratios of $\sim 20$ for gas within the shell and this would apply to B$_{||}$ as well. Since the mean local magnetic field is about 2~$\mu$G \citep[see, e.g.,][]{hml+06}, we can take B$_{||}$ within the shell to be $\sim 40\;\mu$G. Thus the observation of $\Delta$RM $=0.5~\rm rad.~m^{-2}$ implies $\Delta$DM$= 0.016~\rm pc~cm^{-3}$. Such a difference could be due to inhomogeneities because the $\Delta$DM over the 6.5 years between the \citet{sbg+19} observations and our observations is $-0.148$\;cm$^{-3}$\,pc.

We can estimate turbulent $D_{\rm DM}(S_d)$, where $S_d$ is the diffractive scale, from the bandwidth using the technique discussed in \citet{bdz+22}. A bandwidth of 10 MHz at 1375 MHz implies $S_d = 1 \times 10^7$\;m. The phase structure function is unity at $S_d$ and $D_{\rm DM}(S_d) = 2.9 \times 10^{-15}$\;cm$^{-6}$\,pc$^2$. Finally $D_{\rm DM}(12\;{\rm AU}) = 1.6 \times 10^{-6}$\;cm$^{-6}$\,pc$^2$, where have assumed a Kolmogorov spectrum (index +5/3) for the structure function. Thus the rms change in DM over 12~AU is $\Delta$DM$\sim 0.0012$\;cm$^{-3}$\,pc. This is an order of magnitude too small to explain our time variations. Consequently they are unlikely to be turbulent in origin.

There is a significant upward trend in our estimated values of RM$_{\rm ism}$ over the year, from about +22.5~radians~m$^{-2}$ to +23.0~radians~m$^{-2}$. However, we note that \citet{sbg+19} give a value for RM$_{\rm ism}$ of $+22.73\pm 0.08$~rad~m$^{-2}$, approximately the mean value of our observations, measured using the LOFAR high band centered at 150 MHz, on MJD 56747 (2014 March 31). This shows clearly that the fluctuations in B$_{||}$ and/or the electron density in the Monogem Ring have complicated spatial structure on scales of 10's to 1000's of AU. Currently, the only DM measurements sufficiently accurate to study DM fluctuations of order 0.01\;cm$^{-3}$\,pc, as are implied by our $\Delta$RM observations, are the LOFAR observations of \citet{sbg+19}. Although we can't hope to observe the turbulent variations on an AU scale, further LOFAR observations should allow a study of the density inhomogeneities in the Monogem Ring on an AU scale. 


\section{Analysis of RVM fitting}\label{app:rvm}
The process of fitting the RVM model (Equation~\ref{eq:rvm})  
to the observations is not straightforward because of the periodicities and discontinuities in the equation, highly correlated parameters, and non-gaussian errors in the observations \citep[see, e.g.,][]{ew01}. Of the parameters $\alpha$, $\zeta$, $\phi_0$ and $\psi_0$, the one 
that we are most interested in is $\psi_0$, the orientation of the rotational pole projected on the sky. The curves of $\psi(\phi)$ have a reflection symmetry about the point $(\phi=\phi_0, \psi=\psi_0)$.
At this point the gradient $|d\psi/d\phi|$ is maximum and is given by $|d\psi/d\phi|_{\rm max} = \sin(\alpha)/\sin(\beta)$ where $\beta = \zeta - \alpha$.

\begin{figure}[!htp]
\center
 \includegraphics[width=14.0cm, angle=270]{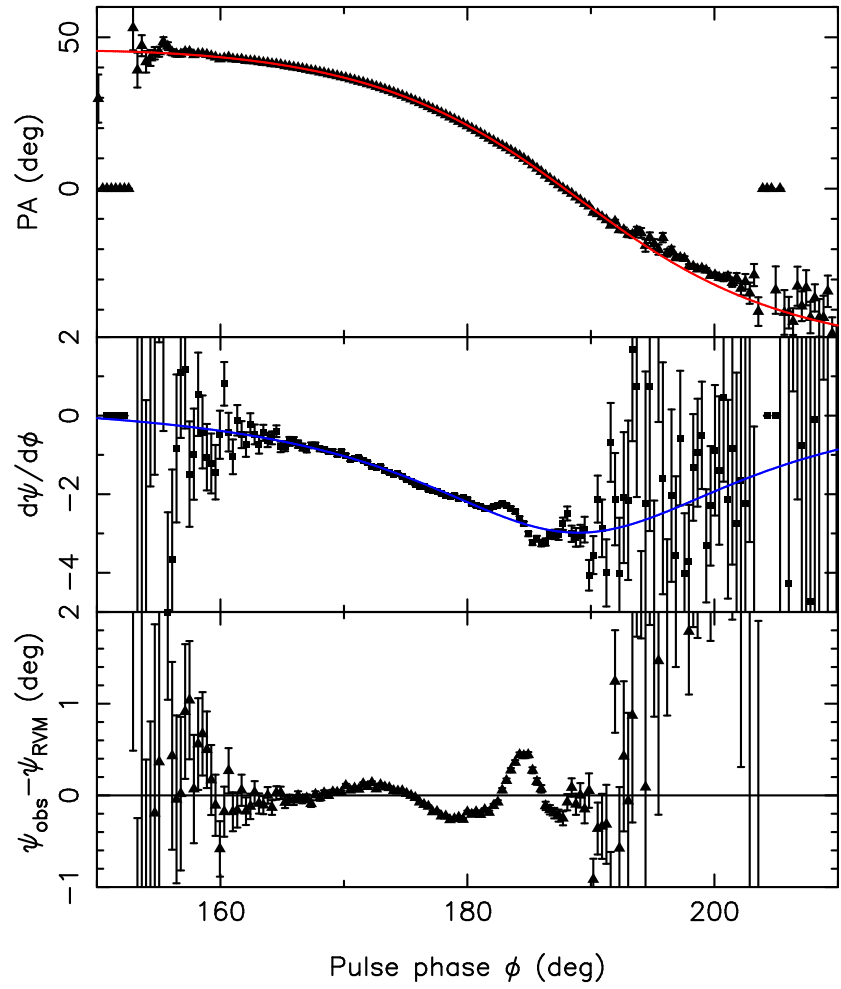}
    \caption{Results of RVM fitting of the observed PAs at 1375~MHz. In the top and middle panels, the red and blue lines show the fitted model. See text for more details. }\label{fig:rvm_fit2}
\end{figure}

An attempt to simply fit Equation~\ref{eq:rvm} to the observations did not give satisfactory results because the parameters are highly correlated and $\alpha$ tends to approach $180\degr$ while $\beta$ approaches zero. The Jacobian diverges at both these limits. To avoid this we examined the gradient $d\psi/d\phi$, shown in the middle panel of Figure~\ref{fig:rvm_fit2}. The maximum is relatively well-determined at $\sim -3.0$. Consequently, we re-parameterized the model, substituting  $|d\psi/d\phi|_{\rm max}$ (MXG) for $\beta$ and fitting the RVM to the gradient. This becomes a 3-parameter fit because $\psi_0$ drops out of the gradient, and it provides a good estimate of MXG and $\phi_0$, but it did not give a useful estimate of $\alpha$. Exploration of the model showed that the gradient is very insensitive to $\alpha$ provided that MXG is kept constant.

We then fitted the RVM to the observations using MXG $=-3.0$ as an initial condition and applying upper and lower bounds $\pm$0.2 about this value. For this fit, the observations were weighted by the measurement error. The algorithm was the {\sc Trust Region Reflection}, as implemented in {\sc SciPy}\footnote{\url{https://scipy.org}}. Because of the complexity of the fit we also performed the fit using the same algorithm, but implemented in Matlab. The results were numerically identical. The residuals to both the PA and gradient fits plotted in Figure~\ref{fig:rvm_fit2} show deviations much larger than the measurement errors, and so we used uniform weighting in a second fit.  The parameters for this second calculation matched those of the first calculation within the uncertainty, but the calculated uncertainties are somewhat larger and, we believe, are more reliable.

The free parameters, ($\alpha$, MXG, $\phi_0$, $\psi_0$), obtained from the RVM-fit at MJD 59512 are highly covariant, but much less so than the set ($\alpha$, $\beta$, $\phi_0$, $\psi_0$). The parameters $\phi_0$ and $\psi_0$ are naturally covariant because $\psi_0 = \psi(\phi_0)$. Also $\phi_0$ can be obtained independently of $\psi_0$ with a gradient fit. So this covariance is much less troublesome. 
The cross-correlation coefficients are as follows,
\begin{align}
C(1,2) = 0.878,\,   C(1,3)=-0.986,\,  C(1,4)&=0.980\\
C(2,3) = -0.928,\,  C(2,4)&=0.950\\
C(3,4)&=-0.997.
 \end{align}
When we correct $\psi_0$ for RM to obtain the intrinsic orientation we obtain $-67\fdg6$ or $+112\fdg4$ which differs substantially from the earlier result from \citet{jbb+07}. We attribute this difference to the much higher signal-to-noise ratio of the FAST observations, which allowed us to fit the RVM through the point of symmetry. \citet{jbb+07} were unable to observe closer than $1-2\degr$ to $\phi_0$ which made their fit much less stable. 

\end{appendix}

\clearpage


\end{document}